\documentclass[twocolumn,showpacs,superscriptaddress,amsmath,amssymb,prl]{revtex4}

\usepackage{graphicx}
\usepackage{bm}

\begin{document}

\title{Strongly inhibited transport of a 1D Bose gas in a lattice}

 \author{C. D. Fertig}
 \affiliation{National Institute of Standards and Technology, Gaithersburg, MD  20899-8424}
 \affiliation{University of Maryland, College Park, MD  20742}
 \author{K. M. O'Hara}
 \altaffiliation{Current address: Department of Physics, Pennsylvania State University, University Park, PA  16802}
 \affiliation{National Institute of Standards and Technology, Gaithersburg, MD  20899-8424}
 \author{J. H. Huckans}
 \affiliation{National Institute of Standards and Technology, Gaithersburg, MD  20899-8424}
 \affiliation{University of Maryland, College Park, MD  20742}
 \author{S. L. Rolston}
 \affiliation{National Institute of Standards and Technology, Gaithersburg, MD  20899-8424}
 \affiliation{University of Maryland, College Park, MD  20742}
 \author{W. D. Phillips}
 \affiliation{National Institute of Standards and Technology, Gaithersburg, MD  20899-8424}
 \affiliation{University of Maryland, College Park, MD  20742}
 \author{J. V. Porto}
 \affiliation{National Institute of Standards and Technology, Gaithersburg, MD  20899-8424}

\date{\today}

\begin{abstract}
We report the observation of strongly damped dipole oscillations of
a quantum degenerate 1D atomic Bose gas in a combined harmonic and
optical lattice potential. Damping is significant for very shallow
axial lattices (0.25 photon recoil energies), and increases
dramatically with increasing lattice depth, such that the gas
becomes nearly immobile for times an order of magnitude longer than
the single-particle tunneling time. Surprisingly, we see no
broadening of the atomic quasimomentum distribution after damped
motion. Recent theoretical work suggests that quantum fluctuations
can strongly damp dipole oscillations of 1D atomic Bose gas,
providing a possible explanation for our observations.

\end{abstract}


\maketitle

The ability of highly degenerate quantum systems to sustain
dissipationless flow is one of the most striking manifestations of
quantum mechanics. However, transport in such systems can be
dramatically modified by the presence of a relatively weak, but
rapidly spatially varying (``corrugated'') potential along the
transport axis. For example, the periodic potential of an optical
lattice inhibits transport in a degenerate Fermi atomic gas
\cite{Mod03, Pez04, Ott04}, but not, in general, in a degenerate
Bose gas (i.e., Bose-Einstein Condensate (BEC)) \cite{Mor01,Cat01}.
However, under certain conditions, highly dissipative transport in a
BEC in an optical lattice \cite{Bur01,Fer02,Cat03} can arise from
nonlinear dynamical instabilities \cite{Wu01, Sme02, Fel04}.  In low
dimensional systems, of which 1D atomic gases
\cite{Tol04,Sto03,Mor03,Par04,Kin04} and superconducting nanowires
\cite{Bez00} are important experimentally realized examples, a
corrugated potential can cause dramatic changes in ground state and
transport properties.

We study inhibited transport in a 1D Bose gas in the presence of an
optical lattice along the 1D axis. In the absence of such a lattice,
dipole oscillations are undamped \cite{Mor03}, since it is a general
result that the dipole mode of a harmonically confined gas is
unaffected by 2-body interactions (generalized Kohn's theorem)
\cite{Dal99}. This result does not strictly hold for a combined
harmonic and periodic potential; nevertheless, undamped oscillations
have been observed in 3D BECs for small amplitudes and weak
interactions\cite{Cat01, Kra02}.

In this Letter we report a study of strongly damped dipole
oscillations of a 1D Bose gas in a combined harmonic and periodic
potential, under conditions for which undamped motion has been
observed previously for 3D BECs.  This striking difference between
1D and 3D was recently reported, qualitatively, in Ref.
\cite{Sto03}. Here we measure the damped motion as a function of
axial lattice depth.  Significant damping is induced by very shallow
lattices, and in deeper lattices the motion is overdamped to the
degree that the gas is nearly immobile for times an order of
magnitude longer than the single-particle tunneling time. We
emphasize, and discuss further below, that the inhibited transport
is not due to Bloch oscillations \cite{Dah96,Mor01}, where transport
is frustrated by Bragg reflection at the Brillouin zone (BZ)
boundary, as has been seen in previous experiments
\cite{Fer02,Mod03,Pez04}.

Our method to realize an ensemble of independent 1D Bose gases is
similar to earlier work \cite{Tol04}. We produce a nearly pure
$^{87}$Rb condensate of $N=(0.8-1.6)\times10^5$ atoms in the $|F=1,
m_F=-1\rangle$ state in a Ioffe-Pritchard magnetic trap
($\nu_x=\nu_z=29$~Hz, $\nu_y=8$~Hz). We next partition the BEC into
an array of independent, vertical 1D ``tubes'' by adiabatically
applying a transverse (in the $xy$ plane) 2D confining lattice
\cite{Gre01,Tol04,Sto03,Mor03,Par04,Kin04}. The confining lattice is
ramped on during $200$~ms to a depth of approximately $30E_R$ (where
$E_R = h^2/2m\lambda^2$ is the photon recoil energy, and $\lambda$
is the laser wavelength).  The combined magnetic and optical
potential results in approximately 5000 occupied tubes, each with an
axial frequency of $\omega_0/2 \pi\approx60$~Hz.  We observe a
Thomas-Fermi density envelope in the combined magnetic and optical
potential, and calculate \cite{Tol04} cloud radii of
$r_x=14(1)~\mu$m \cite{note4}, $r_y=20(1)~\mu$m, and
$r_z=10.6(5)~\mu$m for $N=1.4\times10^5$. From this we estimate a
peak 1D density of $4.8(4) \times10^{4}$~cm$^{-1}$ in the central
tube, and a peak 3D density of
$4.7(4)\times10^{14}~\textrm{cm}^{-3}$. Subsequently, we corrugate
the tubes by adiabatically applying, over $20$~ms, an axial
(vertically along $z$) 1D lattice. The Rayleigh length of the axial
lattice beams is large enough that they do not significantly modify
the axial harmonic potential. All lattice beams derive from a single
Ti:Sapphire laser operating at $\lambda=810$~nm, far detuned from
the atomic resonances at 780 nm and 795 nm.  The pairs of lattice
beams are detuned from each other by 6 MHz, making them effectively
independent \cite{Win99}. The final configuration consists of three
independent standing waves, each formed from a pair of
counter-propagating beams.

\begin{figure}
\includegraphics*{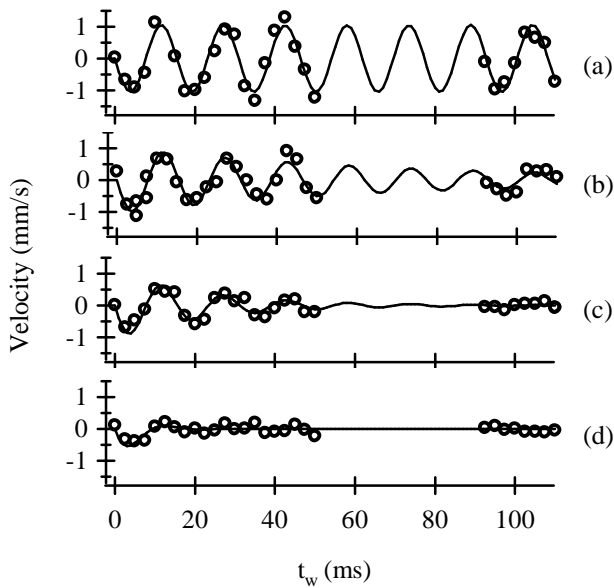}
\caption{\label{fig:underdamped} Damped oscillations of a 1D Bose
gas in an optical lattice.  Shown are plots of velocity versus wait
time $t_w$ from $t_w=0$ to $110$~ms, and for axial lattice depths of
(a) 0$E_R$, (b) $0.25E_R$, (c) $0.50E_R$, and (d) $2.0E_R$, where
$E_R$ is the photon recoil energy (see text).}
\end{figure}

We excite dipole oscillations of the center-of-mass of the atoms in
all the tubes by suddenly ($\lesssim150\mu$s) applying a linear
magnetic field gradient, thus displacing the total harmonic trap
(but not the lattice) axially by $z_0\approx3 \mu$m. This
displacement is less than 30\% of $r_z$, and corresponds to
approximately eight axial lattice sites \cite{Note3}. The waists of
the Gaussian transverse lattice beams ($w_0\approx 210~\mu$m) are
much larger than both $z_0$ and the size of the trapped cloud.

The oscillation in the position of the atoms is too small for our
imaging system to clearly resolve.  We therefore observe oscillation
in \emph{velocity} by waiting a variable time $t_w$ after the
initial displacement, then suddenly turning off all trapping
potentials (with time constants of $\approx250~\mu$s and
$\approx150\mu$s for the optical and magnetic potentials,
respectively), and imaging the atoms after a time-of-flight
$t_{\textrm{TOF}} = 18.4$~ms. The turn-off of the optical lattice is
fast compared to the oscillation period, but slow enough to avoid
diffraction of the atoms (i.e., adiabatic with respect to band
excitations).

We observe damped dipole oscillations for axial lattice depths from
$V=0E_R$ to $2E_R$, as seen in Fig.~\ref{fig:underdamped}. In the
absence of an axial lattice, we observe oscillations (period
$T=15.4$~ms) consistent with no damping (Fig. 1a), indicating that
tube-to-tube dephasing and trap anharmonicities are not significant
on the timescale of our experiments. However, the oscillations are
noticeably damped in a lattice only $0.25E_R$ deep.  Such a shallow
lattice modulates the atomic density by only 6\%, and modifies the
single-particle energy-quasimomentum dispersion relation $E(q)$ from
that of a free-particle around only the last few percent of the BZ.
(We note here, and discuss further below, that the amplitude of
motion is kept well within the quadratic part of $E(q)$ for shallow
lattices).

Beyond a lattice depth of $\approx3E_R$ the motion is overdamped,
and there are no oscillations. In this case, the atoms' velocity can
be quite small, so we use a technique that maps the atoms'
\emph{position} in the trap to the cloud position after TOF. The
experiment proceeds as before, except that after the trap is
displaced by $z_0$, the atoms are allowed to relax toward their
equilibrium position at $z=0$ for a fixed time $t_w=90$~ms. We then
rapidly (with time constant $\approx250~\mu$s) turn off \emph{only
the axial lattice}. The remaining transverse lattice and magnetic
potentials are left on for $3.75$~ms (approximately a quarter-period
of undamped axial harmonic motion), then turned off simultaneously
(as in the underdamped experiment). This converts the axial
displacement $z(t_w)$ into a velocity, which we measure by TOF.

\begin{figure}
\includegraphics*{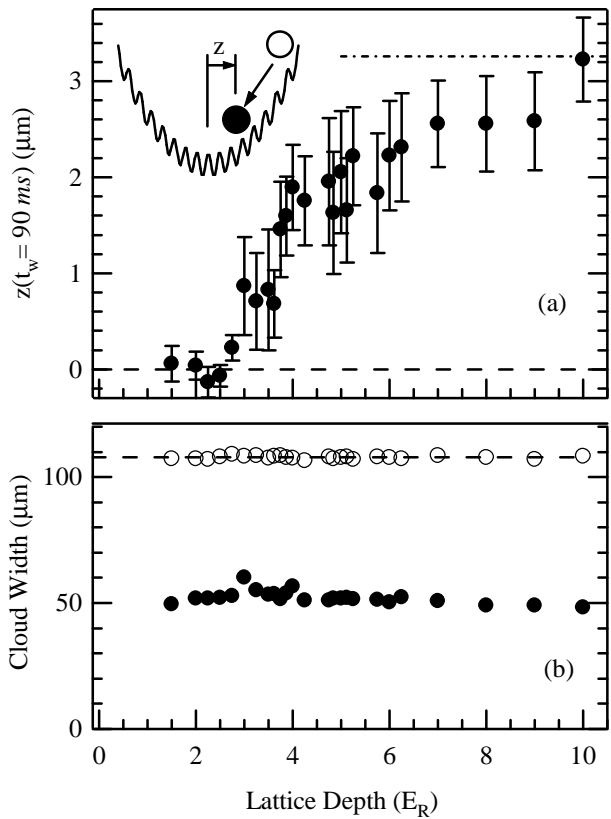}
\caption{\label{fig:overdamped} Overdamped motion of a 1D Bose gas
in an optical lattice. (a) Plot of the atoms' position $90$~ms after
shifting the trap, as a function of lattice depth. Inset depicts
relaxation of the atoms toward equilibrium.  Immediately after the
trap is displaced to $z_0$, the atoms (open circle) begin to move
toward equilibrium, reaching a displacement $z$ from equilibrium
after $90$~ms (solid circle). Also shown is the initial position of
the atoms (dash-dot line). (b) The $1/e$ half-widths of Gaussian
fits to axial TOF distributions (closed symbols), and the
half-widths of square transverse TOF distributions (open symbols)
resulting from a uniformly filled BZ (see text, and
Fig.~\ref{fig:slice}). Also shown (dashed line) is the BZ calculated
from lattice parameters.}
\end{figure}

Figure \ref{fig:overdamped}a shows $z(t_w = 90$~ms) as a function of
axial lattice depth. For the shallowest lattices this wait time is
sufficient for the atoms to damp to the equilibrium position $z=0$.
For the deepest lattices the motion is so overdamped that there is
negligible motion during this time, and the position remains
$z\approx z_0$. We note that in the absence of damping, atoms would
tunnel through the lattice to the equilibrium position in a time
$(T/4)\sqrt{m^*/m}$, where $m^*$ is the effective mass. This time is
only 8~ms for non-interacting particles in a $10E_R$ lattice.

\begin{figure}
\includegraphics*{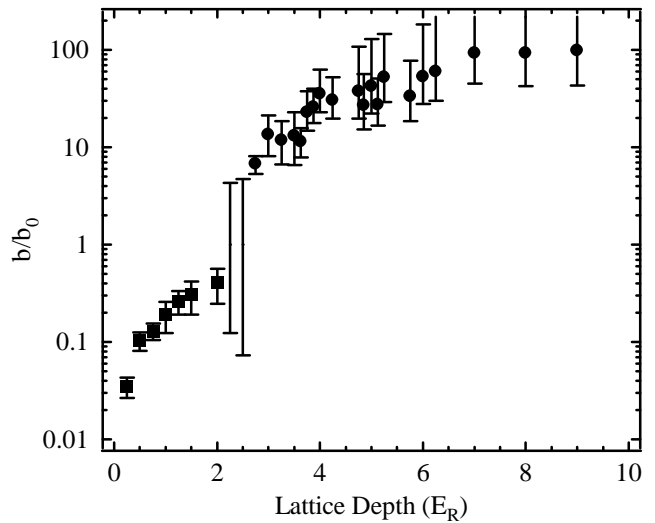}
\caption{\label{fig:dampingparams} Plot of the reduced damping
constant $b/b_0$ for various depths $V$ of the axial lattice, as
determined using the underdamped (squares) and overdamped (circles)
experimental techniques. Near critical damping ($V\approx3E_R$), the
analysis cannot distinguish between underdamped and overdamped
motion, so only upper and lower bounds are shown at $V=2.25$ and
$2.50$.}
\end{figure}

To quantify the damping we model the motion as damped simple
harmonic, $m^* \ddot{z}= -b \dot{z} -kz$, and extract a damping
constant $b=b(V)$ for different axial lattice depths $V$. For
underdamped motion, we simultaneously fit the oscillation data for 8
depths to the expression
$$\dot{z}(t) = A \frac{k}{\omega m^*} e^{-bt/2m^*} \sin(\omega t),$$
where $\omega\equiv\sqrt{k/m^*-(b/2m^*)^2},$ and $A$ and $k$ are fit
parameters common across all $V$. For overdamped motion, we
determine $b=b(V)$ from the overdamped solution for $z(t_w)$, which
in the limit of strong damping simplifies to
$$z(t_w)\approx z_0 \left( \frac{e^{-k t_w/b}}{1-km^*/b^2}+\frac{e^{-b t_w/m^*}}{1-b^2/km^*} \right),$$
where $z_0$ and $k$ are inputs derived from measurements of undamped
oscillations.  In our analysis we use a single-particle calculation
of the effective mass $m^*$ \cite{Note2}.

Figure \ref{fig:dampingparams} shows a plot of $b(V)/b_0$ versus
lattice depth V, where $b_0\equiv2m\omega_0$ corresponds to
critically damped harmonic motion for $\omega_0\equiv\sqrt{k/m}$. We
show data for both the under- and overdamped regimes, and note that
the damping constant increases by at least a factor 1000 for a
30-fold increase in lattice depth.

\begin{figure}
\includegraphics*{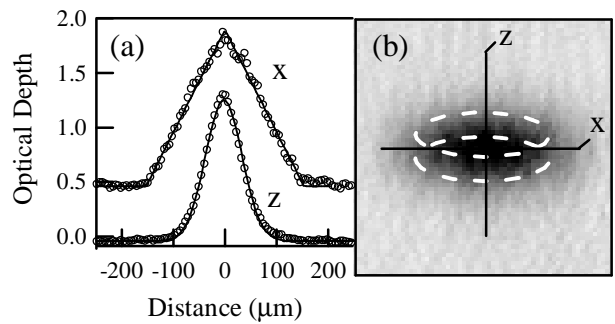}
\caption{\label{fig:slice} Cross-sections (a) of a TOF absorption
image (b) of the expanded atom cloud. The plot of the transverse
cross-section (along $x$) is offset vertically by 0.5 units for
clarity.  The solid lines are Gaussian (triangular) fits to the
axial (transverse) profiles, respectively. The dashed ovals in (b)
indicate the peak-to-peak range of motion of undamped dipole
oscillations.}
\end{figure}

The axial width of the cloud after TOF can provide information about
the distribution of atomic quasimomenta in the lattice. The lattice
turn-off time constant of $250~\mu$s is long enough to avoid
diffraction, but short enough to be non-adiabatic with respect to
inter-well tunneling and interactions. (Related experiments in 2D
\cite{Gre01} and 3D \cite{Gre02} lattices support this conclusion.)
In the absence of interactions, and neglecting the initial size of
the cloud, the turn-off maps the single-particle quasimomentum
distribution of atoms in the lattice to free particle (i.e.,
plane-wave) momentum states that can be directly observed in TOF. In
the presence of interactions, the mapping is complicated by
mean-field repulsion during TOF.   A variational calculation
\cite{Rey04} indicates that mean-field repulsion is in fact the
dominant contributor to the axial TOF width in our system.
Therefore, the extracted TOF width greatly overestimates the width
of a narrow initial quasimomentum distribution.

An example TOF image is shown in Fig.~\ref{fig:slice}, together with
cross-sectional profiles of the optical depth along the axial and
transverse directions. The first BZ of the transverse lattice is
uniformly filled, producing a uniform, square spatial distribution
(in the $xy$ plane) after TOF.  Our imaging system views this square
distribution along the diagonal in the $xy$ plane, resulting in a
triangular profile, from which we extract the width of the square
(open symbols, Fig.~\ref{fig:overdamped}b). In the axial direction
the distribution is narrower and reasonably well fit to a Gaussian,
from which we extract the axial (along $z$) TOF $1/e$ half-width
(filled symbols, Fig.~\ref{fig:overdamped}b). Even for the strongly
overdamped data, the axial TOF width (which, we recall, overstates
the width of narrow quasimomentum distributions), is much narrower
than the BZ. This implies that the inhibition of transport is not
due to effects related to Bloch oscillations of a filled BZ, as
observed in Refs. \cite{Mod03,Fer02}.  Furthermore, we do not see a
significant difference in TOF width between atoms that undergo
damped harmonic motion and those that are unexcited but held for an
equal time \cite{note1}.  This is in stark contrast to earlier
experiments on 3D BECs \cite{Bur01,Cat03,Fel04}, where strong
damping was accompanied by a pronounced broadening and fragmentation
of the quasimomentum distribution.

Large amplitude dipole oscillations in a lattice can damp due to
dynamical instabilities caused by particle interactions. For a 3D
BEC moving in an optical lattice, such an instability point occurs
at $q \geq q_\pi/2$ ($q_\pi\equiv2\pi/\lambda$ is the BZ boundary),
where the dispersion relation has an inflection point, as predicted
in \cite{Wu01,Sme02}, and observed in \cite{Bur01,Cat03,Fel04}. This
effect is manifested as a large increase in the width of the
quasimomentum distribution. Here, in contrast, we keep the maximum
(single-particle) quasimomentum $q_{\textrm{max}}$ of the
oscillation small by limiting the initial energy of displacement
$E(z_0)=m \omega_0^2 z_0^2/2$.  For $V<2E_R$, our choice of $z_0$
corresponds to $q_{\textrm{max}}\approx q_\pi/5$. For deeper
lattices our fixed $z_0$ corresponds to a larger $q_{\textrm{max}}$,
but is always less than $q_\pi/2$ for $V<9E_R$.  (In a separate
experiment, we excited oscillations in our system with twice the
usual amplitude, and saw stronger damping that was accompanied by a
broadening of the axial TOF width by nearly a factor of two.)

For small amplitude dipole oscillations  ($q_{\textrm{max}}\ll
q_\pi/2$) of a 3D BEC in an optical lattice, the effect of the
lattice is merely to increase the effective mass, leading to
undamped motion at a lower frequency \cite{Cat01, Kra02}.  In the
reduced dimensionality system of our 1D Bose gas, we have seen that
the optical lattice has a qualitatively different effect.  To
highlight the difference between these two situations, we excited
dipole oscillations in a 3D BEC (i.e., \emph{no} transverse
confining lattice) in a $4E_R$ axial lattice, and saw no damping.
This is in contrast to the results of the same experiment in a 1D
Bose gas (i.e., \emph{with} a transverse confining lattice), shown
in Fig.~\ref{fig:dampingparams}, where $b(V=4E_R)/b_0\approx50$
corresponds to extremely overdamped motion.

After we performed these experiments, theoretical treatments
appeared which suggested that zero-temperature quantum fluctuations
can lead to substantial damping of transport in a 1D atomic Bose gas
\cite{Pol04, Ban04}.  Our observations, including the significant
damping in lattices too shallow to support a Mott-Insulator (MI)
phase \cite{Kol04}, can be explained by the mechanisms of Refs.
\cite{Pol04, Ban04}, but appear to be inconsistent with a mechanism
involving incompressibility \cite{Sto03}.

It is possible that there is a temperature dependence to the
damping; unfortunately, we can derive little information on
temperature from the TOF widths. Future experiments to investigate
the temperature dependence could shed light on the relative
importance of quantum and thermal fluctuations to dissipation, a
question of interest in, for example, the development of ultrathin
superconducting wires \cite{Bez00}. We also look forward to testing
other explicit predictions of these theories, such as the dependence
of damping on displacement and dimensionality. We note that the
periodic potential of an optical lattice is free from defects, and
that 1D atomic Bose gases are well isolated from the environment,
yielding a relatively clean system in which to compare experiment
with theory. The ability to continuously and dynamically vary the
confining potentials makes optical lattice experiments attractive
for future studies of superfluidity in low dimensional quantum
systems.

The authors gratefully acknowledge helpful discussions with
Ana-Mar{\'i}a Rey, Julio Gea-Banacloche, and Guido Pupillo. This
work was partially supported by ARDA.

\bibliographystyle{apsrev}
\bibliography{Fertig111904}

\begin{thebibliography}{31}
\bibitem[{not()}]{note4}
\bibinfo{note}{Error bars here and everywhere represent the uncorrelated
  combination of 1 standard deviation statistical and systematic uncertainties
  (if any).}
\bibitem[{not()}]{note1}
\bibinfo{note}{During the $\sim 100$~ms in which we observe the cloud's motion,
  we observe a small growth in width of the TOF image, at a rate that is
  independent of excitation or damping, consistent with heating due to
  spontaneous scattering of lattice photons.}
\end{thebibliography}
\end{document}